\documentclass[12pt]{article}
\usepackage{amssymb,amsmath,epsfig}

\begin{document}
\title{\bf Shearfree Spherically Symmetric Fluid Models}

\author{M. Sharif \thanks{msharif.math@pu.edu.pk} and Z. Yousaf
\thanks{z.yousaf.math@live.com}\\
Department of Mathematics, University of the Punjab,\\
Quaid-e-Azam Campus, Lahore-54590, Pakistan.}

\date{}

\maketitle
\begin{abstract}
We try to find some exact analytical models of spherically
symmetric spacetime of collapsing fluid under shearfree condition.
We consider two types of solutions: one is to impose a condition
on the mass function while the other is to restrict the pressure.
We obtain totally of five exact models, and some of them satisfy
the Darmois conditions.
\end{abstract}

Gravitational collapse and compact body evolution under various
conditions are important in general relativity. Their description
can be found by exploring the dynamical equations of spherically
symmetric models. Generally, this requires kinematics of such
fluids having expansion, acceleration, rotation and shear or
distortion. There have been many interesting results arising due
to shearfree conditions. The vanishing of shear tensor describes
the physical aspects of compact bodies in the relativistic
astrophysics phenomena.

Collins and Wainwright$^{[1]}$ reported that the class of
shearfree expanding (or collapsing) irrotational perfect fluids
with an equation of state $p=p(\mu)$ is either FRW or spherical
symmetric Wyman solution or a special case of plane symmetric
models. Misra and Srivastava$^{[2]}$ explored that charged perfect
fluids with vanishing shear tensor and uniform density are
necessarily static. Tomimura and Nunes$^{[3]}$ discussed a
radiating spherical body collapse with heat flow having the
property of shearfree and the geodesic motion of the fluid.

Glass$^{[4]}$ showed that the shearfree perfect fluid is
irrotational and also stationary vacuum spacetime is also static
if and only if the Weyl tensor is of purely electric type. Carr
and Coley$^{[5]}$ found that every shearfree perfect fluid
solution is self-similar but converse is not true. Herrera {\it et
al}.$^{[6]}$ studied the stability of the shearfree condition of a
spherically symmetric local anisotropic fluid with null radiation,
shearing viscosity and dissipation in the form of heat flux.
In$^{[7]}$, Di Prisco {\it et al}. found some exact analytical
models of spherically symmetric spacetime under the expansion free
condition. In this Letter, we extend this work to investigate some
exact analytical models under the shearfree condition.


The interior region is given by the most general spherically
symmetric metric
\begin{equation}\label{1}
ds^2_-=-A^2(t,r)dt^{2}+B^2(t,r)dr^{2}+R^2(t,r)(d\theta^{2}
+\sin^2\theta{d\phi^2}).
\end{equation}
We assume comoving coordinates inside the hypersurface
${\it\Sigma}$. The energy-momentum tensor is of the form
\begin{equation}\label{2}
T^-_{\alpha\beta}=(\mu+P_{\bot})V_{\alpha} V_{\beta}+P_\bot
g_{\alpha\beta}+{\it\Pi} \chi_{\alpha} \chi_{\beta},
\end{equation}
where ${\it\Pi} \equiv P_r - P_\bot$, and $\mu,~P_{\perp},~P_r$,
$V^{\alpha},~\chi^{\alpha}$ are the energy density, the tangential
and radial pressure, the four-velocity and a unit four-vector
along the radial direction respectively. They satisfy
\begin{equation*}
V^{\alpha}V_{\alpha}=-1,\quad\chi^{\alpha}\chi_{\alpha}=1,\quad
\chi^{\alpha}V_{\alpha}=0.
\end{equation*}
The expansion scalar, ${\it\Theta}$, four acceleration,
$a_{\alpha}$ and shear tensor $\sigma_{\alpha\beta}$ read
\begin{eqnarray*}\label{4}
{\it\Theta}=V^{\alpha}_{;\alpha},\quad
a_\alpha=V_{\alpha;\beta}V^\beta,\quad
\sigma_{\alpha\beta}=V_{(\alpha;b)}+a_{(\alpha} V_{\beta)}
-\frac{1}{3}{\it\Theta}(g_{\alpha\beta}+V_\alpha V_\beta).
\end{eqnarray*}
The non-vanishing components of the shear tensor are
\begin{eqnarray*}\label{5}
\sigma_{11}=\frac{2}{3}B^2\sigma,\quad
\sigma_{22}=\frac{\sigma_{33}}{\sin^2\theta}=-\frac{1}{3}R^2\sigma,
\end{eqnarray*}
where
\begin{equation}\label{6}
\sigma=\frac{1}{A}\left(\frac{\dot{B}}{B}
-\frac{\dot{R}}{R}\right).
\end{equation}
The four-acceleration gives
\begin{eqnarray*}\label{7}
&a_{1}=\frac{A'}{A},\quad a^2=a^\alpha
a_{\alpha}=(\frac{A'}{AB})^2, \quad a^{\alpha}=a\chi^{\alpha},
\end{eqnarray*}
\begin{equation}\label{8}
{\it\Theta}=\frac{1}{A}\left(\frac{\dot{B}}{B}
+2\frac{\dot{R}}{R}\right),
\end{equation}
where dot and prime stand differentiation with respect to $t$ and
$r$, respectively.

The Einstein field equations for the interior spacetime
are$^{[7]}$
\begin{eqnarray}\nonumber
8{\pi}{\mu}A^{2}&=&\left(\frac{2\dot{B}}{B}
+\frac{\dot{R}}{R}\right)\frac{\dot{R}}{R}\\\label{10}&-&\left(\frac{A}{B}\right)^2
\left[\frac{2R''}{R}+\left(\frac{R'}{R}\right)^2-\frac{2B'R'}{BR}
-\left(\frac{B}{R}\right)^2\right],\\\label{11}
0&=&-2\left(\frac{\dot{R'}}{R}-\frac{\dot{B}A'}{BR}
-\frac{\dot{R}A'}{RA}\right), \\\nonumber
8{\pi}P_{r}B^{2}&=&-\left(\frac{B}{A}\right)^2\left
[\frac{2\ddot{R}}{R}-\left(\frac{2\dot{A}}{A}
-\frac{\dot{R}}{R}\right) \frac{\dot{R}}{R}\right]\\\label{12}
&+&\left(\frac{2A'}{A}+\frac{R'}{R}\right)\frac{R'}{R}
-\left(\frac{B}{R}\right)^2, \\\nonumber
8{\pi}P_{\perp}R^{2}&=&8{\pi}P_{\perp}R^{2}\sin^{-2}\theta\\\nonumber
&=&-\left(\frac{R}{A}\right)^2\left[\frac{\ddot{B}}{B}
+\frac{\ddot{R}}{R}-\frac{\dot{A}}{A}
\left(\frac{\dot{B}}{B}+\frac{\dot{R}}{R}\right)
+\frac{\dot{B}\dot{R}}{BR}\right]\\\label{13}
&+&\left(\frac{R}{B}\right)^2\left[\frac{A''}{A}
+\frac{R''}{R}-\frac{A'B'}{AB}\right.
\left.+\left(\frac{A'}{A}-\frac{B'}{B}\right)\frac{R'}{R}\right].
\end{eqnarray}
The mass function is given as$^{[8]}$
\begin{equation}\label{14}
m(t,r)=\frac{R}{2}(1-g^{\alpha\beta}R_{,\alpha}R_{,\beta})
=\frac{R}{2}\left(1+\frac{\dot{R}^2}{A^2} -\frac{R'^2}{B^2}\right).
\end{equation}
Using the velocity of the collapsing fluid $U=\frac{\dot{R}}{A}$ in
Eq.(\ref{14}), we have
\begin{equation}\label{20}
E\equiv\frac{R'}{B}=\left[1+U^{2}-\frac{2m(t,r)}{R}\right]^{1/2}.
\end{equation}
It follows from Eq.(\ref{14}) that
\begin{equation}\label{15}
\dot{\mu}R'+{P_r}'\dot{R}+(P_r+\mu)\dot{(R'}+2R'\frac{\dot{R}}{R})=0.
\end{equation}
The conservation of the energy-momentum tensor leads to
\begin{eqnarray}\nonumber
\dot{\mu}+A\sigma(\mu+P_r)+3(\mu+P_\perp)\frac{\dot{R}}{R}+{\it\Pi}\frac{\dot{R}}{R}=0,\\\label{28}
{P_{r}}'+(\mu+P_r)\frac{A'}{A}+2{\it\Pi}\frac{R'}{R}=0.
\end{eqnarray}
When we take the Schwarzschild metric outside 3D hypersurface
${\it\Sigma}$ as the exterior spacetime, we can write by using
junction conditions$^{[7]}$
\begin{eqnarray}\label{47}
Adt\overset{{\it\Sigma}}=dv\left(1-2\frac{M}{\rho}\right),\quad
R\overset{{\it\Sigma}}=\rho(v),\quad
m(t,r)\overset{{\it\Sigma}}=M,\quad P_{r}\overset{{\it\Sigma}}=0,
\end{eqnarray}


Next, we take the shearfree fluid and explore some exact
analytical models. We would like to mention here that the
shearfree fluids make our analysis and results of purely local
character. Under this condition, i.e., $\sigma=0$, Eq. \,(\ref{6})
turns out to be $\frac{\dot{B}}{B}=\frac{\dot{R}}{R}$, which gives
\begin{equation}\label{48}
B={\gamma}R,
\end{equation}
where $\gamma$ is an arbitrary function of $r$ which is taken as $1$
without loss of generality. Using this value of $B$ in
Eq.(\ref{11}), we obtain
\begin{equation}\label{49}
A=\frac{\dot{R}}{R\xi},
\end{equation}
where $\xi$ is an arbitrary function of $t$. The physical
variables $\mu,~P_r$ and ${\it\Pi}$ can be written in terms of $R$
and $m$ as
\begin{eqnarray}\label{50}
4\pi\mu=\frac{m'}{R'R^{2}},\quad
4\pi{P_{r}}=-\frac{\dot{m}}{\dot{R}R^{2}},\quad
{\it\Pi}=-\left[R\frac{\dot{\mu}}{\dot{R}}+3(\mu+P_\bot)\right].
\end{eqnarray}
Using Eqs.(\ref{14}), (\ref{48}) and (\ref{49}), it follows
\begin{equation}\label{51}
m(t,r)=\frac{R}{2}\left(R^{2}\xi^{2}-\frac{R'^{2}}{R^{2}}+1\right).
\end{equation}
We can see that the metric coefficients $A$ and $B$ are now given
interms of $R$. In the following, we obtain some exact analytical
models.

In this family of solutions, we assume $m(t,r)$ as follows$^{[9]}$
\begin{equation}\label{52}
2m(t,r)=jR+\frac{1}{3}kR^{3}+\frac{1}{5}lR^{5},
\end{equation}
where $j,~k$ and $l$ are arbitrary functions of $t$. The energy
density can be obtained by using Eqs.(\ref{50}) and (\ref{52}) as
\begin{equation}\label{53}
8\pi\mu=\frac{j}{R^{2}}+k+lR^{2}.
\end{equation}
Using Eqs.(\ref{50})-(\ref{52}), it follows that
\begin{equation}\label{54}
j-1+kR^{2}+lR^{4}-3(R\xi)^{2}+2\frac{R''}{R}-\left(\frac{R'}{R}\right)^{2}=0.
\end{equation}
Equations (\ref{51}), (\ref{52}) and (\ref{54}) give
\begin{equation*}
-2(j-1)+\frac{2}{5}lR^{4}-4\left(\frac{R'}{R}\right)^{2}+2\frac{R''}{R}=0.
\end{equation*}
We assume ${R^{2}}\equiv{S}$ so that the above equation
implies
\begin{equation*}
aS+bS^{3}-\frac{3}{2}\frac{S'^2}{S}+S''=0,
\end{equation*}
where $a(t)\equiv-2(j-1),~b(t)\equiv\frac{2}{5}l$. Integrating
this equation, we have
\begin{equation}\label{57}
S'^2={2}\left(-aS^{2}+bS^{4}\right).
\end{equation}
We would like to solve this equation for the following cases:

{\it Case} (i) $a\neq0,~b\neq0$. Integration of Eq.(\ref{57})
yields
\begin{equation}\label{58}
S=\sqrt{\frac{a}{b}}\sec\left(\sqrt{2a}(r+\beta)\right),
\end{equation}
or
\begin{equation}\label{59}
R=\frac{5(1-j)}{l}\left[\sec\left(2\sqrt{1-j}(r+\beta)\right)\right]^\frac{1}{2},
\end{equation}
where $\beta(t)$ is an arbitrary function. Consequently,
Eq.(\ref{50}) yields
\begin{eqnarray}\nonumber
8\pi\mu&=&\frac{j}{R^{2}}+lR^{2}, \\\nonumber
8\pi{P_r}&=&-\left(\frac{j+lR^{4}}{R^{2}}\right)
-\frac{R^{3}}{\dot{R}}\left(\frac{\dot{l}}{5}+\frac{\dot{j}}{R^{4}}\right),\\\label{62}
8\pi{P_{\bot}}&=&3lR^{2}+\frac{3j}{R^{2}}+\frac{3R^{3}\dot{l}}{5\dot{R}},
\end{eqnarray}
Using junction conditions (\ref{47}), we obtain two independent
equations with three unknown functions $j(t),~l(t)$ and $\beta(t)$
which can be satisfied by any convenient choice of one of these
functions. This does not lead to interesting solutions.

{\it Case} (ii) $a=0$. This case gives $j=1$ and Eq.(\ref{57})
leads to
\begin{equation}\label{63}
S=\left(\frac{1}{r{\sqrt{2b}}+\beta(t)}\right), \quad or\quad
R=\left(\frac{\sqrt{5}}{2r{\sqrt{l}}+\sqrt{5}\beta(t)}\right)^\frac{1}{2}.
\end{equation}
The physical variables turn out to be
\begin{eqnarray}\nonumber
8\pi\mu&=&\frac{2r\sqrt{l}+\sqrt{5}{\beta}}{\sqrt{5}}
+\frac{l{\sqrt{5}}}{{2r\sqrt{l}+\sqrt{5}{\beta}}},\\\nonumber
8\pi{P_r}&=&-\left(\frac{2r\sqrt{l}+\sqrt{5}{\beta}}{\sqrt{5}}
+\frac{l{\sqrt{5}}}{{2r\sqrt{l}+\sqrt{5}{\beta}}}+\frac{2\dot{l}\sqrt{l}}{r\dot{l}
+\sqrt{5l}\dot{\beta}}\right),\\\label{62}
8\pi{P_{\bot}}&=&-\frac{2l{\sqrt{5}}}{{2r\sqrt{l}
+\sqrt{5}{\beta}}}+\frac{4\dot{l}\sqrt{l}}{r\dot{l}+\sqrt{5l}\dot{\beta}}.
\end{eqnarray}
{\it Case} (iii) $b=0$. Here we obtain $l=0$ while integration of
Eq.(\ref{57}) yields
\begin{equation}\label{68}
S=e^{\left(r{\sqrt{2a}}+\beta(t)\right)},\quad or\quad
R=e^{\left(r{\sqrt{(1-j)}}+\frac{\beta(t)}{2}\right)}.
\end{equation}
The quantities $\mu,~P_r$ and $P_{\bot}$ take the form
\begin{eqnarray}\nonumber
8\pi\mu&=&j\left[e^{\left(2r\sqrt{1-j}+\beta\right)}\right]^{-1},\quad
{P_{\bot}}=0.\\
8\pi{P_r}&=&-\left[e^{\left(2r\sqrt{1-j}+\beta\right)}\right]^{-1}
\left[\dot{j}\left(\frac{-r\dot{j}}{2\sqrt{1-j}+\frac{\dot{\beta}}{2}}\right)^{-1}+j\right].\label{62}
\end{eqnarray}

Lemaitre$^{[10]}$ firstly used fluid spheres with tangential
stresses alone ($P_{r}=0$). Afterwards, the use of this kind of
fluid is in practice by many authors$^{[11-17 ]}$. The second family
of solution will be obtained by assuming $P_{r}=0$. Using this
condition in Eq.(\ref{15}), it follows that
\begin{equation}\label{73}
R^{3}=3\int\frac{C_{1}(r)}{\mu}dr+C_{2}(t),
\end{equation}
where $C_{1}$ and $C_{2}$ are arbitrary functions. Using $P_{r}=0$
in Eq.(\ref{15}), we have $\mu=C_1(r)/R'R^2$. Comparing this value
with that in Eq.(\ref{50}), we obtain $C_{1}=m'/4\pi$. Using the
shearfree condition in the first equation of continuity, we have
\begin{equation}\label{75}
P_{\bot}=-\left(\frac{\dot{\mu}{R}}{2\dot{R}}+\frac{3{\mu}}{2}\right).
\end{equation}
Now we shall explore different models for some particular cases.

{\it Case (i)} 1: $P_{\bot}=\alpha\mu$. For $P_{r}=0$ and
$P_{\bot}=\alpha\mu$, the second of Eq.(\ref{28}) yields
\begin{equation}\label{76}
\dot{R}=f(t)R^{(2{\alpha}+1)},\quad R'=g(r)R^{(2{\alpha}+1)},
\end{equation}
or
\begin{equation}\label{77}
R^{-2\alpha}=\psi(t)+\chi(r),
\end{equation}
where
$\psi(t)={(-2{\alpha})}\int{f(t)}dt,~\chi(r)={(-2{\alpha})}\int{g(r)}dr$
while $f(t)$ and $g(r)$ are arbitrary functions. Without loss of
generality, we can choose $f(t)=\xi(t)$, then Eqs.(\ref{49}) and
(\ref{76}) yield $A=R^{2\alpha}$. Using Eqs.(\ref{14}), (\ref{48})
and (\ref{76}) on the hypersurface $r=r_{e}$ with $\gamma=1$, it
follows that
\begin{equation*}\label{81}
\dot{R}^{2}\overset{{\it\Sigma}}{=}R^{4{\alpha}}\left(\frac{2M}{R}+{g^{2}}{R^{4\alpha}}-1\right),
\end{equation*}
which can be solved for an arbitrary value of $\alpha$. We choose
$M=1$. When $\alpha=1/4$, we obtain
\begin{equation}\label{82}
R\overset{{\it\Sigma}}{=}\frac{1}{\sqrt{2}g}\left[1+\tan{\sqrt{2}{(t+t_{0})}}\right],
\end{equation}
and hence Eq.(\ref{77}) yields
\begin{equation}\label{83}
\psi(t)\overset{{\it\Sigma}}{=}\left[\frac{1}{\sqrt{2}g}\left(1+
\tan{\sqrt{2}{(t+t_{0})}}\right)\right]^\frac{-1}{2}+\chi.
\end{equation}
Thus the time dependence of all variables is fully determined. Now
the radial dependence ($C_1$ or $\chi$) can be obtained from the
initial data.

{\it Case (ii)} 2: ${\mu}={\mu_{0}}C_{1}/r^{2}$. Here we assume
that energy density is separable so that
${\mu}={\mu_{0}(t)}C_{1}/r^{2}$. Consequently, Eqs.(\ref{73}) and
(\ref{75}) give
\begin{eqnarray}\label{85}
R=\left(\frac{r^{3}}{\mu_{0}}+C_{2}(t)\right)^{1/3}, \quad
P_{\bot}=-\frac{3{\mu_{0}}^2C_{1}}{2{r^2}}\left[\frac{C_{2}\dot{\mu_{0}}
+\dot{C_{2}}\mu_{0}}{\dot{C_{2}}{\mu_{0}}^2-\dot{\mu_{0}}r^3}\right].
\end{eqnarray}
Equation (\ref{51}) provides the mass function
\begin{equation}\label{87}
m=\frac{R}{2}\left[R^{2}\xi^{2}-\frac{r^{4}R^{-6}}{{\mu_{0}^{2}}}+1\right].
\end{equation}
If we take $\xi R=\dot{R}$, then $
A_{{\it\Sigma}}=1,~\dot{R_{{\it\Sigma}}}=U_{{\it\Sigma}},~
\xi=\frac{U_{{\it\Sigma}}}{R_{{\it\Sigma}}}$. Using these values
along with Eq.(\ref{47}) in (\ref{87}), it follows that
\begin{equation}\label{89}
R_{{\it\Sigma}}=2M\left[U_{{\it\Sigma}}^{2}-\frac{r_{{\it\Sigma}}^{4}
\xi^{6}}{{\mu_{0}^{2}}U_{{\it\Sigma}}^{6}}+1\right]^{-1}.
\end{equation}
Thus in the absence of superluminal velocities $(U<1)$ and
Eq.(\ref{89}), we must impose $r_{{\it\Sigma}}^{2}<{\mu_{0}}$. We
can find the time dependence of $R_{\it\Sigma}$ if
$\mu_{0}=\mu_{0}(t)$, which implies time dependence of all
variables.


We have found two families of solutions by imposing conditions on
mass function and pressure. The first family yields three while
the second gives two exact analytical models. The solutions of the
second family satisfies the junction conditions which contain some
of the essential features of a realistic situation. These models
may be helpful for the analysis of gravitational behavior of
compact bodies. The solutions may not show any specific
astrophysical scenario but they describe the possible importance
of the shearfree condition for exploring exact models. Also, this
work may be considered as a toy model of localized systems. We
would like to mention here energy density changes with time even
under the shearfree condition which is obvious from equation of
continuity.

\end{document}